\documentclass[aps, prl, amsmath, preprintnumbers, superscriptaddress, twocolumn, amssymb, floatfix]{revtex4-1}
\usepackage{graphicx}
\usepackage{mathptmx, textcomp}
\usepackage[latin1]{inputenc}

\begin{document}
\title{Long-term drifts of stray electric fields in a Paul trap}
\author{Arne H\"arter}
\affiliation{Institut f\"ur Quantenmaterie and Center for Integrated Quantum Science and Technology IQ$^\textrm{ST}$,
Universit\"at Ulm, 89069 Ulm, Germany}

\author{Artjom Kr\"ukow}
\affiliation{Institut f\"ur Quantenmaterie and Center for Integrated Quantum Science and Technology IQ$^\textrm{ST}$,
Universit\"at Ulm, 89069 Ulm, Germany}

\author{Andreas Brunner}
 \altaffiliation[Present address:]{3. Physikalisches Institut, Universit\"at Stuttgart, 70569 Stuttgart, Germany}
 \affiliation{Institut f\"ur Quantenmaterie and Center for Integrated Quantum Science and Technology IQ$^\textrm{ST}$,
Universit\"at Ulm, 89069 Ulm, Germany}

\author{Johannes Hecker Denschlag}
\affiliation{Institut f\"ur Quantenmaterie and Center for Integrated Quantum Science and Technology IQ$^\textrm{ST}$,
Universit\"at Ulm, 89069 Ulm, Germany}
             
\begin{abstract}
We investigate the evolution of quasi-static stray electric fields
in a linear Paul trap over a period of several months. Depending
on how these electric fields are initially induced we
observe very different time scales for the field drifts.
Photo-induced electric fields decay on time scales of days. We
interpret this as photo-electrically generated charges on insulating
materials which decay via discharge currents. In
contrast, stray fields due to the exposure of the ion trap to a
beam of Ba atoms mainly exhibit slow dynamics on the order
of months. We explain this observation as a consequence of a coating
of the trap electrodes by the atomic beam. This may lead to
contact potentials which can slowly drift over time due to
atomic diffusion and chemical processes on the surface. In order not to
perturb the field evolutions, we suppress the generation of
additional charges and atomic coatings in the Paul trap during the
measurements. For this, we shield the ion trap from ambient light
and only allow the use of near-infrared lasers. Furthermore, we
minimize the flux of atoms into the ion trap chamber. Long-term
operation of our shielded trap led us to a regime of very low
residual electric field drifts of less than 0.03$\,$V/m per day.
\end{abstract}

\maketitle

\section{Introduction}
\label{intro}
Paul traps have become essential tools in widely different fields of research ranging
from quantum information$\,$\cite{Lei03,Bla08,Haf08} and quantum simulation$\,$\cite{Bla12} to
precision metrology$\,$\cite{Ros08} and cold collisions between ions and
neutrals$\,$\cite{Smi10,Zip10,Vul2008,Hud2011,Ran2012,Hall2011}. The further development
of all these lines of research hinges on continuing improvements of the Paul trap architectures
and on a better understanding of the current experimental issues. \\
Ideally, a single ion in a Paul trap is only subjected to the
electric fields generated by the voltages applied to the trap
electrodes. However, even small spatial variations of the
electrode surface potential (i.e. patch potentials) in the
vicinity of the trap center create stray electric fields which
significantly perturb this ideal configuration. This leads to
undesired experimental complications. Quasi-static stray fields
lead to positional shifts of the trapped
ion$\,$\cite{Har10,Wan11,Daniilidis2011,Yu1991,DeVoe2002,Nar11,All11b,Deb08}
and thus to excess micromotion$\,$\cite{Ber98}.
Rapidly fluctuating stray fields lead to ion heating
$\,$\cite{Yu1991,DeVoe2002,Nar11,All11b,Tur00,Des06,Lab08,All11,Hit12}.

For the following discussion, we will formally group
patch potentials into two categories.\\
Patch potentials of category 1 can decay via electronic discharge
currents, similar to a capacitor that is shorted with a resistor.
Thus, their evolution is in general governed by the motion of
electrons. For example, the photoelectric effect can generate
charges on the dielectric surfaces of the Paul trap. These
surfaces could be the trap mounts but also insulating oxide layers
on the trap electrodes. The photoelectric effect typically appears
with light at wavelengths below about 500$\,$nm. As most of the
trapped ion species require light at such ``blue" wavelengths for
laser cooling and interrogation, light-induced patch charges will
continuously be created while the experiments are carried out. In
addition, patch potentials of category 1 can also be generated in
other ways such as the direct deposition of electrons and ions on
dielectric surfaces (see e.g. \cite{Daniilidis2011}).\\
Patch potentials of category 2 are stable as long as the surface
atoms do not move. They are due to a spatial variation of the
material$'$s work function which depends on its composition, its
crystal orientation and its surface adsorbates \cite{Opa92,Bro92}.
Surface adsorbates can be elements or compounds which are
physically or chemically bound onto the surface. Crystal
orientation comes into play for neighboring grains in a
polycrystalline structure. A change in composition comes about
e.g. when two different metals are brought into contact. This
gives rise to the contact potential, i.e. the difference of the
work functions of the two metals. Ion traps are often loaded from
atomic beams which are directed towards the trap center. If the
atomic beam hits the Paul trap electrodes, atoms are deposited on
the electrode surface, potentially forming contact potentials.
Furthermore, as previously mentioned, the formation and deposition
of chemical compounds on the trap surface can also create
electric stray fields.\\
Patch potentials of both categories have been observed to cause
deteriorations of Paul traps leading to strong ion heating
effects$\,$\cite{Yu1991,DeVoe2002,Nar11,Tur00,Des06,Lab08,All11,Hit12}.
However, the influence and evolution of the patch potentials of
category 2 was often masked by the presence of
photo-induced patch potentials.\\
Here, we study the long-term dynamics of quasi-static electric fields in a
Paul trap in an environment where we systematically suppress both continuous
surface contamination and continuous photo-induced patch
charge build-up. We observe smooth drifts of the quasi-static electric stray
fields on various time scales from days to months. We interpret these
different time scales as indications of different physical and chemical
processes that take place. For example, electric fields induced by the
exposure of the trap to laser light typically decay within a few days. Fields
induced by creating an atomic beam using a barium oven show slower
dynamics on the order of months. After longer
time periods without surface contamination and photo-induced
charging, the stray electric fields settle smoothly towards a
stable value with very small residual drifts as low as 0.03$\,$V/m
per day.\\
In our setup, we achieve the suppression of surface contamination
and photo-induced patch charges as follows. We create and probe
ions (Rb$^+$) in a linear Paul trap by using only near-infrared
light sources ($\lambda$ = 780$\,$nm and 1064$\,$nm) and small clouds of
$\approx 10^5$ ultracold Rb atoms, which have previously been
optically transported into the chamber$\,$\cite{Har13}. Thus, the net
flux of atoms into the chamber is negligible. We measure the
electric stray fields by applying
compensating electric fields until the excess micromotion of the ion is
minimized$\,$\cite{Har13b}.

\section{Experimental setup and methods}
\label{sec:1} \label{sec:2}

\begin{figure}
\resizebox{0.48\textwidth}{!}{%
  \includegraphics{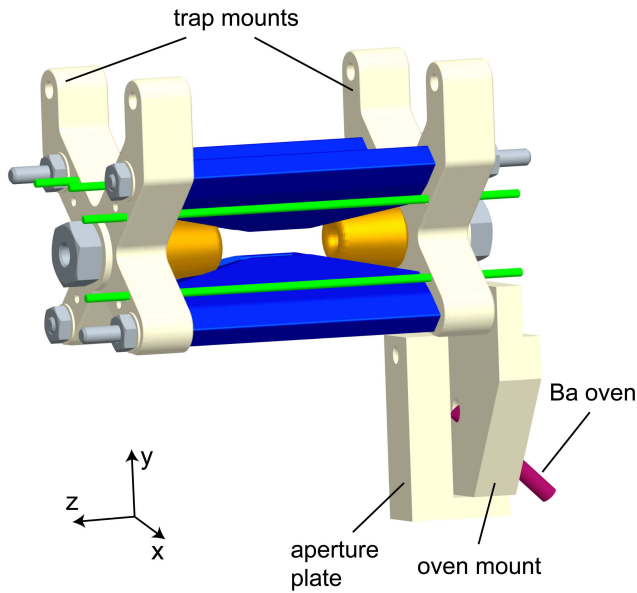}
}
\caption{Paul trap with mounts. The trap consists of four rf electrodes (blue),
two endcap electrodes (yellow) and two pairs of compensation electrodes (green).
The mounts for the trap and the barium oven are made of MACOR. The aperture plate
was installed to reduce the amount of barium deposited on the trap electrodes.
}
\label{fig:1}
\end{figure}

The design of our linear Paul trap is shown in
Fig.$\,$\ref{fig:1}. The effective distance from the trap center
to each of the four radiofrequency (rf) electrodes is 2.6$\,$mm
while the distance to the endcap electrodes measures $7\,$mm. To
create radial confinement, a voltage driven at a frequency of
$4.17\:$MHz with an amplitude of $500\:$V is applied to two of the
rf electrodes while the other two are held at ground potential.
Axial confinement is generated by applying static voltages of
about $8\:$V to the two endcap electrodes. Under these
experimental conditions, a $^{87}$Rb$^+$ ion is confined at radial
trapping frequencies of about $350\:$kHz and an axial trapping
frequency of about $50\:$kHz. The total depth of the trap is on
the order of 4$\,$eV and allows for ion storage times of many
days, even without any type of cooling. The Paul trap is part of a
hybrid atom-ion trap setup that brings the trapped ion into contact
with an ultracold cloud of atoms$\,$\cite{Smi12}. Ensembles of
$^{87}$Rb atoms are prepared in a separate vacuum
chamber and transported into the Paul trap using a long-distance
optical transport line. They are then loaded into a crossed dipole
trap where further evaporative cooling down to typical
temperatures of 700$\,$nK is performed. The atom numbers typically
range between $10^5$ and $10^6$ atoms. Both the optical transport
and the crossed dipole trap are implemented using several W of
laser power at a wavelength of 1064$\,$nm. To perform absorption
imaging of the atoms, resonant laser light at 780$\,$nm is used.
After this destructive imaging process, a new atom cloud is
prepared within 30$\,$s for the next measurement. To load an ion,
a Rb atom cloud with a density of several
$10^{13}\,\textrm{cm}^{-3}$ is prepared and positioned at the
center of the Paul trap. Three-body recombination processes in the
atom cloud produce Rb$_2$ molecules which are subsequently ionized
by a REMPI process using photons from the dipole trap
laser$\,$\cite{Har13}. Subsequently, the molecular Rb$_2^+$ ions
quickly dissociate via collisions with neutral atoms and finally
yield Rb$^+$ ions.\\
To detect the number of trapped Rb$^+$ ions and measure their micromotion,
we employ a sensitive probing scheme using ultracold atomic
clouds$\,$\cite{Har13b}. For this, we immerse the ions into clouds
consisting of about $10^5$ atoms at densities around
$10^{12}\,\textrm{cm}^{-3}$. After a few seconds of interaction
time, we detect the final atom number and atom temperature which
depend on the number of ions and their micromotion. This enables us
to reliably work with a single ion and, by minimizing its excess
micromotion with the help of electric compensation fields, to
measure the stray electric fields acting on the
ion$\,$\cite{Ber98,Har13b}. Due to the production time required
for the atom clouds, an electric field measurement requires about an
hour of measurement time, resulting in a limited temporal
resolution. The measurement precision is high and typically ranges
around 0.1$\,$V/m for the results presented here.

\begin{figure}
\resizebox{0.48\textwidth}{!}{%
  \includegraphics{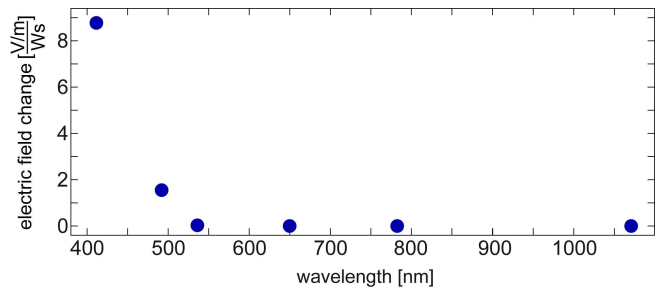}
}
\caption{Changes of the vertical stray electric field normalized to laser power and exposure time.
The field changes were observed to be linear in both laser power and exposure time.
At wavelengths below 500$\,$nm, light-induced electric fields sharply increase.
}
\label{fig:2}
\end{figure}

\section{Results}
\label{sec:3}

As a first step we investigate the susceptibility of our trap
setup with respect to laser light at various wavelengths which are available in
our lab (Fig.$\,$\ref{fig:2}). For these and the following measurements, each laser beam
propagates through the center of the ion trap in horizontal
direction at an angle of 45$\,^\circ$ with respect to the trap
axis. In particular, the laser beam and its specular reflection
from the vacuum windows do not directly impinge on any trap electrode or
trap mount. Only some stray scattered light from the windows
illuminates the trap parts diffusely and quite evenly. Both the
trap electrodes and the trap mounts can be ``charged up"
via the photoelectric effect. The trap mounts and the mounts
for the barium oven are made of machinable glass-ceramic (MACOR), which is
very susceptible for accumulating charges. Charges can also
accumulate on the trap electrode surfaces as these often feature
undesired insulating coatings such as oxide layers. As expected, for
wavelengths below 500$\:$nm we find a sharp increase in
light-induced build-up of electric fields (see Fig.2). For
780$\,$nm or the even longer wavelength of 1064$\,$nm, we did not
detect any measurable light-induced electric fields.

In general, we find the photo-induced stray electric fields to be pointing mostly
in vertical (+$y$-axis) and axial (+$z$-axis) direction. There might be a number
of reasons how this asymmetry in the direction of the electric stray field comes
about. One reason could be that
the laser light illuminates the trap setup asymmetrically.
However, we can exclude this possibility as the field direction is quite insensitive
to changes in the propagation direction of the laser light, in particular when it
is flipped by 90$^\circ$ in the horizontal plane. Another possible explanation for
the asymmetry could be the presence, e.g. of a single dust
particle on the trap electrodes, which is something we cannot
rule out. An obvious asymmetry, however, is already
inherent in our Paul trap setup due to the location of the mount
and aperture plate of the barium oven (see Fig.$\,$1, lower right). Indeed,
we estimate that these two parts (which are made of MACOR) can lead to
significant electric stray fields at the location of the ion.
For one, they can potentially be charged to voltages of up to 500$\,$V,
as determined by the amplitude of the rf trap drive. (The rf fields prevent the
patch potentials from saturating at a low voltage, the value of which would be
normally set by the difference of the work function and the photon energy.)
Secondly, our trap geometry is relatively open such that external electric fields
can penetrate quite well to the position of the ion. We cannot
deterministically charge the aperture plate or the oven mount to a certain
voltage in order to test their effect on the electric field at the
position of the ion. However, we can apply voltages to the Ba oven itself
which should create electric fields of similar magnitude and orientation.
We find that a voltage of 1$\,$V on the oven indeed results in a dominant
electric field contribution of $+0.1\,$V/m in $y$-direction. The other
two field components are each about a factor of 6 smaller ($-0.016\,$V/m
in $x$-direction and $+0.016\,$V/m in $z$-direction). These values set a lower
bound to the expected fields originating from a charged aperture plate
which is located closer to the ion than the oven.

\begin{figure}
\resizebox{0.49\textwidth}{!}{%
  \includegraphics{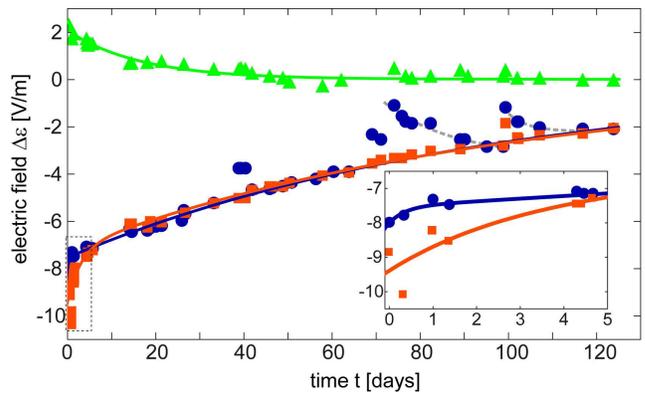}
} \caption{Long-term drift of the horizontal (x-direction, orange
squares), vertical (y-direction, blue circles) and axial
(z-direction, green triangles) electric fields. Except for two
occasions ($t\approx 70\:$days and $t\approx 100\:$days), the trap
was isolated from any light below a wavelength of 780$\:$nm. Solid
lines are double-exponential fits with long-term time constants on
the order of three months. Offsets of the electric fields are
chosen such that $\Delta \varepsilon$ converges towards 0 in the
long-term limit. \textit{Inset}: Zoom into the initial
field evolution with time constants of 0.3 and 2.7 days (cf. table 1).}
\label{fig:3}
\end{figure}

Next, we start a long-term experiment where we monitor the
evolution of all three spatial components of the stray electric
field over a time span of about four months (Fig.$\,$\ref{fig:3}).
Before the start of these measurements,
the ion trap was operated with Ba$^+$ ions so that both the barium
oven and the necessary lasers were frequently used. Consequently,
there is a substantial stray electric field to begin with. 
During the measurements (except for two short occasions) the whole
experimental setup is almost entirely shielded from ambient light
by means of light-tight protective covers to avoid any patch
charge build-up. All three electric field components
(in $x$, $y$, $z$ direction) show a more or less monotonic decay and
converge towards long-term limits which are each set to
$\Delta \varepsilon = 0$ in the plot. The solid lines in Fig.$\,$\ref{fig:3}
are double-exponential fits of the form
\begin{center}
$\varepsilon(t)=\Delta \varepsilon_1 \exp(-t/\tau_1) + \Delta
\varepsilon_2 \exp(-t/\tau_2) ,$
\end{center}
where $\Delta \varepsilon_{1,2}$ are the electric field shifts and
$\tau_{1,2}$ are the time constants of the exponential decay
curves.
For the two radial directions
($x$- and $y$-directions) we find relatively rapid initial decays with
time constants $\tau_1= 0.3-2.7$ days (see inset of Fig.$\,$\ref{fig:3}) and
subsequent slow decays with $\tau_2\approx 90$ days. In axial direction the
time constants are $0.6$ and $18\;$days. For all three directions the
slow decays are dominant as they account for roughly 80-95$\,$\%
of the electric field shifts (see table$\,$1).\\
On two occasions ($t\approx 70\;$days and $t\approx 100\;$days), the
light-tight protective covers around the experimental setup had to be removed
for several hours so that the Paul trap was subjected (quite uniformly) to ambient
white light from the fluorescent ceiling lights. As a consequence,
the electric field in vertical ($y$) direction shows a sharp
increase and then decays back towards its long-term behavior
within several days. We investigate this effect in detail below.\\
After about 100 days, the daily drift of the vertical field was
below $0.03\:$V/m yielding extremely stable experimental
conditions. In addition, this slow drift allows for a precise
prediction of the expected electric fields at a given time so that
the field compensation can be adjusted without requiring
additional measurements.

\begin{figure}
\resizebox{0.48\textwidth}{!}{%
  \includegraphics{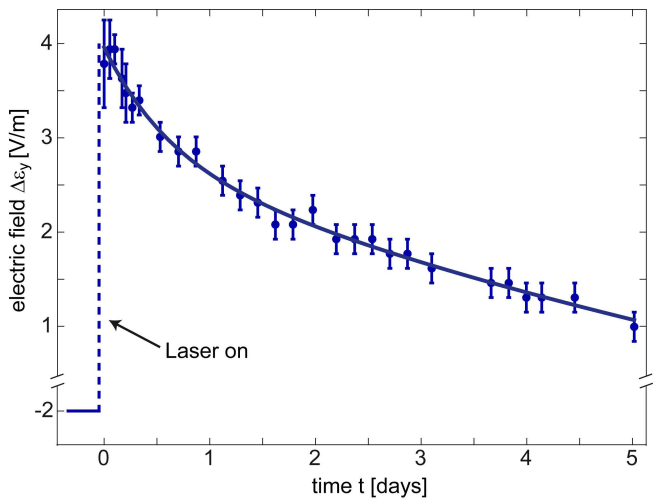}
}
\caption{Vertical electric field shift induced by subjecting the Paul trap to light at 413$\:$nm. The subsequent
relaxation is fit by a double-exponential function with a dominant slow decay accounting for 80\% of the
field shift. The corresponding time constant is $\tau_2= 11\:$days.}
\label{fig:4}
\end{figure}

After the time period shown in Fig.$\,$\ref{fig:3}, we make use of
the low stray field drift to selectively test the
dynamics of photo-induced patch charges. For this measurement we
shine through the chamber about 2.5$\,$mW of laser power at a
wavelength of 413$\,$nm (3$\,$eV) for 4 minutes.
The direction of the laser beam and the conditions with respect to trap illumination
are the same as for the measurements in Fig.$\,$2, as discussed in the beginning of section 3.
The strongest effect is again observed in the vertical electric field component which
increases by about 6$\:$V/m. The axial field component increases
by 1.5$\,$V/m, the horizontal component by 0.6$\,$V/m. The laser
was then switched off and the decay of the vertical field
component was monitored over 5 days (Fig.$\,$\ref{fig:4}). The
data is fit by a double-exponential curve with an initial decay on
a timescale $\tau_1 = 1.2\:$days and a slow decay with $\tau_2 =
11\:$days. This slow decay accounts for about 80$\,$\% of the
field shift. The observed initial increase and subsequent decay of
the electric field are in rough agreement with the behavior seen
in Fig.$\,$\ref{fig:3} after the Paul trap had been subjected to
ambient light. Thus, our data in Figs. 3 and 4 clearly indicate that
photo-induced electric fields decay on typical timescales of a
few days.\\
We explain the evolution of the photo-induced stray
fields as follows: Initially, the photoelectric effect generates
charges on electrically isolated surfaces resulting in electric
stray potentials. Over time these potentials discharge by
the finite resistivity of the material. In case of a
charging of the aperture
plate or the oven mount we can estimate a timescale for
the discharge. At room temperature, MACOR has a specific volume
resistivity of about 10$^{17} \Omega\times$cm. The typical
resistance of a cm-sized component then is on the order of
10$^{17} \Omega$. The electric capacitance of a cm$^2$-sized plate
is about 4$\epsilon_0\,$cm, with $\epsilon_0$ denoting the vacuum
permittivity. Thus, the decay constant is on the order of 10
hours, which roughly agrees with our observed time scales. We note
that the resistivity of MACOR is strongly temperature dependent.
We do not have a precise knowledge of the temperature of the mount
but from our estimate it is clear that time scales for the
discharging of the MACOR parts should not be much longer than a
few days. The fact that we observe not only a single timescale
for the field decay indicates that there is more than one
contribution to the stray fields. A charged up MACOR part
might exhibit a different discharge behavior than a dust particle
or an isolated patch on the electrode surface. In any case, the
data in Fig. 4 and especially in Fig. 3 (at t $\approx 70$ and t
$\approx 100$ days) clearly indicate that photo-induced
fields always decay on a timescale faster than two weeks. The
question is then how the very slow drift taking place over about
90 days can be explained (see Fig.$\,$\ref{fig:3}). Our observatios
suggest that the electric fields linked to this slow decay have a
different origin than the photoelectric effect.\\
We conjecture that the slow drifts originate from the
dynamics of potentials of category 2, e.g. contact potentials. 
These potentials may change due to slow chemical reactions in the
ultrahigh vacuum environment or by diffusion and migration
processes on the electrode surface. It is known that barium reacts
and forms compounds with O$_2$, N$_2$, CO$_2$ and H$_2$O. It also
acts as a getter material, inclosing non-reactive gases. Thus, the
mobility of Ba on a surface is sizeable. (Interestingly, the work
function of barium is known to remain quite constant ($\approx
2.5\,$eV) even when contaminated with other substances.) Diffusion
or migration of barium on the electrode surface can
coat certain compound layers and set other ones free -- thus
giving rise to slowly changing contact potentials. At room
temperature the vapor pressure of barium is very low. Even at
200$^\circ$C it only reaches $10^{-12}\,$mbar. This suggests that
barium coatings have a long lifetime at ambient temperatures.
\begin{figure}
\resizebox{0.48\textwidth}{!}{%
  \includegraphics{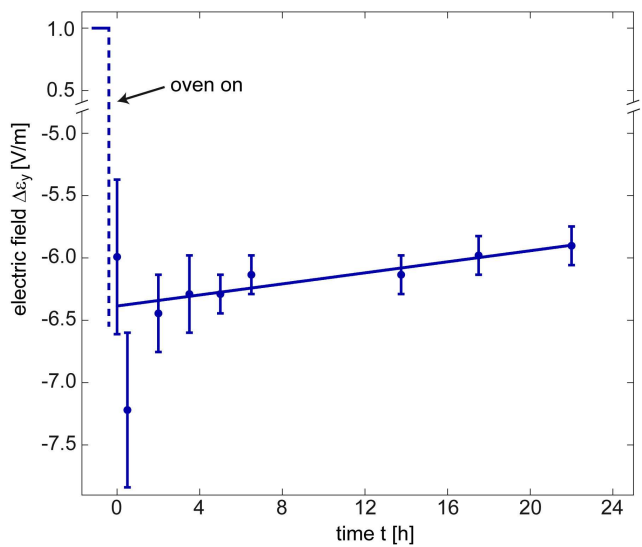}
}
\caption{Vertical electric field shift induced by heating the barium oven for 10 minutes.
During the oven heating time the field dropped by about 7.5$\,$V/m.
After the oven was turned off, the field shows a drift towards higher values.
The straight solid line is a guide to the eye.}
\label{fig:5}
\end{figure}
\begin{table*}
\centering \caption{Overview of the observed drift time constants
and the corresponding electric field shifts. Oven-induced and
light-induced effects give rise to drifts with opposite signs.}
\label{tab:1}
\begin{tabular}{lccccc}
\hline\noalign{\smallskip}
direction & cause & $\tau_1$ [days] & $\Delta \varepsilon_1$[V/m] & $\tau_2$ [days] & $\Delta \varepsilon_2$[V/m]\\
\noalign{\smallskip}\hline\noalign{\smallskip}
vertical (Fig.$\,$\ref{fig:3}) & blue light + oven operation & 0.3$\pm$0.4 & -0.4$\pm$0.2 & 90$\pm$10 & -7.4$\pm$0.4 \\
horizontal (Fig.$\,$\ref{fig:3}) & blue light + oven operation & 2.7$\pm$0.8 & -2.1$\pm$0.3 & 94$\pm$21 & -7.3$\pm$0.7 \\
axial (Fig.$\,$\ref{fig:3}) & blue light + oven operation & 0.6$\pm$0.8 & 0.4$\pm$0.2 & 18$\pm$3 & 1.9$\pm$0.1\\
vertical (Fig.$\,$\ref{fig:4}) & blue light & 1.2$\pm$0.1 & 1.2$\pm$0.1 & 11$\pm$0.3 & 4.9$\pm$0.1 \\
\end{tabular}
\end{table*}
Previous studies have investigated the influence of barium contaminations
on Paul trap electrodes made of Be-Cu$\,$\cite{DeVoe2002}. Long time scales
for the quasi-static electric stray field drifts on the order of months
were found, similar to our time constant for the slow drift ($\tau_2 = 90\:$days)
in Fig.$\,$\ref{fig:3}. Furthermore, it was found in$\,$\cite{Yu1991}
that baking out a Paul trap with tungsten electrodes that had been
coated with Ba significantly changed electric stray fields.

In order to test our conjecture, we now investigate the influence
of the barium oven (while keeping any light below a wavelength of
780$\,$nm blocked off). Immediately after the measurements of Fig.$\,$4
are completed, the oven is heated to a temperature of more than
600$\,^\circ$C for 10 minutes inducing a drop of the vertical
electric field component by about 7.5$\:$V/m
(Fig.$\,$\ref{fig:5}). The horizontal and axial fields each
drop by about 2$\,$V/m (not shown).\\
This electric field drop may be partially explained by a rapid
discharging of left-over light-induced charges on the oven mount
caused by the heat dissipated by the oven. The conductivity of
MACOR increases by more than ten orders of magnitude when heated
from room temperature to several hundred $^\circ$C. It can thus be
expected that any left-over charges in the vicinity of the oven
will be efficiently removed at such high temperatures. However,
the electric field $\Delta \varepsilon$ drops much further than
the initial value of $-2\,$V/m in Fig.$\,$4, namely down to
$-6.5\,$V/m. This additional negative electric field drop cannot
be explained by the discharge of the charges that were previously
produced photoelectrically.\\
One possible explanation for this field contribution
is the effect of contact potentials on the trap electrodes.
Despite the collimation of the atomic beam through the aperture plate,
a fraction of the atoms emerging from the
barium oven reaches the rf electrodes close to the trap center.
Such a coating of the electrodes might immediately change the distribution
of the contact potentials in close vicinity to the ion. These potentials
can be on the order of a few V, as determined by the difference in work
function of the metals involved. In our setup there is a clear
asymmetry on how the Ba oven may coat the trap electrodes with Ba
atoms. The two lower rf electrodes will each be coated only on
one side, whereas the two upper electrodes will probably be coated
over the full tip. It is then to be expected that contact
potentials of the upper electrodes dominate over the potentials of the
lower ones. As the work function of Ba is lower than that of
stainless steel, an electric field component should build up which points towards the
negative $y$-axis. This is indeed what we observe. Furthermore,
there is also an asymmetry in $z$-direction, as the atomic beam from
the oven passes at an angle of about 45$^\circ$ through the blades
(see Fig.$\,$1). This may give rise to an electric stray field
component in $z$-direction.\\
Fig.$\,$5 shows that after the oven is
turned off, the vertical electric field component
increases by about 0.5$\,$V/m per day.  Such a
behavior agrees with the observations at the beginning of the
long-term measurements shown in the inset of Fig.$\,$\ref{fig:3}.
This again supports our interpretation that the 90-day long drift
behavior of the electric field in Fig.$\,$\ref{fig:3} is a result
of initially coating the trap electrodes and mounts with Ba which afterwards
migrates and undergoes chemical reactions on the surface.

\section{Conclusion}
\label{sec:4}

In conclusion, we have investigated the long-term drifts of
quasi-static electric stray fields in a linear Paul trap. We find drifts
on time scales ranging from about half a day to three
months. We suggest that these different time scales reflect different physical or
chemical processes. Light-induced electric fields decay on relatively
short time scales on the order of a few days. This is most probably
due to charges located on insulating material which slowly discharge via
the high electric resistance. In contrast, electric fields which are induced
by turning on the Ba oven exhibit long-term drifts on time scales of up to 90 days.
Guided by analysis, our interpretation is that the oven coats the
trap electrodes or mounts. As a consequence, contact potentials
appear which give rise to the electric fields. These fields show long-term
drifts possibly due to slow migration or reaction processes
taking place on the electrode surface.\\
Patch potentials have been identified as the common source
of both quasi-static and fluctuating electric fields.
The time scales for the drifts and fluctuations of these fields
is reflected by the dynamics taking place on different length scales
of the patches. It has been found that atomic contamination of Paul traps
lead to both quasi-static and rapidly fluctuating patch
fields$\,$\cite{Daniilidis2011,Nar11,DeVoe2002}. Our findings on the
evolution and the origins of quasi-static charges may thus provide
new insights into mechanisms connected to rapidly fluctuating charges
such as anomalous heating effects. The results
presented here are a first investigation with our setup in the
direction of surface dynamics of patch potentials. In the future
the experiments can easily be refined to obtain more detailed
information and to test hypotheses. For example, by locally
applying laser fields on trap mounts and electrodes (either to
heat them up or to produce photo-induced patch charges in a
controlled way), we should be able to spatially probe surface
properties. Another result of our work is that by systematically
avoiding the creation of electric patch potentials we are able to
get into a regime of very small and predictable electric stray
field drifts as low as $0.03\:$V/m per day. Such stability of the
trap conditions may prove valuable for the future development of
precision ion trap experiments. 

The authors would like to thank Stefan Schmid and Wolfgang
Schnitzler for help in the lab. This work was supported by
the German Research Foundation DFG within the SFB/TRR21.

\bibliographystyle{unsrt}

\begin{thebibliography}{10}

\bibitem{Lei03}
D.~Leibfried, R.~Blatt, C.~Monroe, and D.~J. Wineland.
\newblock Rev. Mod. Phys., 75, 281, 2003.

\bibitem{Bla08}
R.~Blatt and D.~Wineland.
\newblock Nature, 453, 1008--1015, 2008.

\bibitem{Haf08}
H.~H\"affner, C.~Roos, and R.~Blatt.
\newblock Phys. Rep., 469, 155, 2008.

\bibitem{Bla12}
R.~Blatt and C.~F. Roos.
\newblock Nature Physics, 8, 277, 2012.

\bibitem{Ros08}
T.~Rosenband, D.~B. Hume, P.~O. Schmidt, C.~W. Chou, A.~Brusch, L.~Lorini,
  W.~H. Oskay, R.~E. Drullinger, T.~M. Fortier, J.~E. Stalnaker, S.~A. Diddams,
  W.~C. Swann, N.~R. Newbury, W.~M. Itano, D.~J. Wineland, and J.~C. Bergquist.
\newblock Science, 319(5871), 1808--1812, 2008.

\bibitem{Smi10}
S.~Schmid, A.~H\"arter, and J.~Hecker~Denschlag.
\newblock Phys. Rev. Lett., 105, 133202, 2010.

\bibitem{Zip10}
C.~Zipkes, S.~Palzer, C.~Sias, and M.~K\"{o}hl.
\newblock Nature, 464, 388, 2010.

\bibitem{Vul2008}
A.~Grier, M.~Cetina, F.~Orucevic, and V.~Vuletic.
\newblock Phys. Rev. Lett., 102, 223201, 2009.

\bibitem{Hud2011}
W.G. Rellergert, S.T. Sullivan, S.~Kotochigova, A.~Petrov, K.~Chen, S.J.
  Schowalter, and E.R. Hudson.
\newblock Phys. Rev. Lett., 107, 243201, 2011.

\bibitem{Ran2012}
K.~Ravi, S.~Lee, A.~Sharma, G.~Werth, and S.~A. Rangwala.
\newblock Nat. Commun., 3, 1126, 2012.

\bibitem{Hall2011}
F.H.J. Hall, M.~Aymar, N.~Bouloufa-Maafa, O.~Dulieu, and S.~Willitsch.
\newblock Phys. Rev. Lett., 107, 243202, 2011.

\bibitem{Har10}
M.~Harlander, M.~Brownnutt, W.~H\"ansel, and R.~Blatt.
\newblock New Journal of Physics, 12, 093035, 2010.

\bibitem{Wan11}
Shannon~X. Wang, Guang~Hao Low, Nathan~S. Lachenmyer, Yufei Ge, Peter~F.
  Herskind, and Isaac~L. Chuang.
\newblock Journal of Applied Physics, 110, 104901, 2011.

\bibitem{Daniilidis2011}
N.~Daniilidis, S.~Narayanan, S.~A. M\"oller, R.~Clark, T.~E. Lee, P.~J. Leek,
  A.~Wallraff, St. Schulz, F.~Schmidt-Kaler, and H.~H\"affner.
\newblock New Journal of Physics, 13(1), 013032, 2011.

\bibitem{Yu1991}
N.~Yu, W.~Nagourney, and H.~Dehmelt.
\newblock Journal of Applied Physics, 69(6), 3779--3781, 1991.

\bibitem{DeVoe2002}
R.~G. DeVoe and C.~Kurtsiefer.
\newblock Phys. Rev. A, 65, 063407, 2002.

\bibitem{Nar11}
S.~Narayanan, N.~Daniilidis, S.~A. M\"oller, R.~Clark, F.~Ziesel, K.~Singer,
  F.~Schmidt-Kaler, and H.~H\"affner.
\newblock Journal of Applied Physics, 110(11), 114909, 2011.

\bibitem{All11b}
D.T.C. Allcock, T.P. Harty, H.A. Janacek, N.M. Linke, C.J. Ballance, A.M.
  Steane, D.M. Lucas, Jr. Jarecki, R.L., S.D. Habermehl, M.G. Blain, D.~Stick,
  and D.L. Moehring.
\newblock Applied Physics B, 107(4), 913--919, 2012.

\bibitem{Deb08}
M.~Debatin, M.~Kr\"oner, J.~Mikosch, S.~Trippel, N.~Morrison, M.~Reetz-Lamour,
  P.~Woias, R.~Wester, and M.~Weidem\"uller.
\newblock Phys. Rev. A, 77, 033422, 2008.

\bibitem{Ber98}
D.~J. Berkeland, J.~D. Miller, J.~C. Bergquist, W.~M. Itano, and D.~J.
  Wineland.
\newblock Journal of Applied Physics, 83(10), 5025--5033, 1998.

\bibitem{Tur00}
Q.~A. Turchette, D.~Kielpinski, B.~E. King, D.~Leibfried, D.~M. Meekhof, C.~J.
  Myatt, M.~A. Rowe, C.~A. Sackett, C.~S. Wood, W.~M. Itano, C.~Monroe, and
  D.~J. Wineland.
\newblock Phys. Rev. A, 61, 063418, 2000.

\bibitem{Des06}
L.~Deslauriers, S.~Olmschenk, D.~Stick, W.~K. Hensinger, J.~Sterk, and
  C.~Monroe.
\newblock Phys. Rev. Lett., 97, 103007, 2006.

\bibitem{Lab08}
J.~Labaziewicz, Y.~Ge, D.R. Leibrandt, S.X. Wang, R.~Shewmon, and I.L. Chuang.
\newblock Phys. Rev. Lett., 101, 180602, 2008.

\bibitem{All11}
D.T.C. Allcock, L.~Guidoni, T.P. Harty, C.J. Ballance, M.G. Blain, A.M. Steane,
  and D.M. Lucas.
\newblock New Journal of Physics, 13(12), 123023, 2011.

\bibitem{Hit12}
D.~A. Hite, Y.~Colombe, A.~C. Wilson, K.~R. Brown, U.~Warring, R.~J\"ordens,
  J.~D. Jost, K.~S. McKay, D.~P. Pappas, D.~Leibfried, and D.~J. Wineland.
\newblock Phys. Rev. Lett., 109, 103001, 2012.

\bibitem{Opa92}
F.~Rossi and G.I. Opat.
\newblock J. Phys. D: Appl. Phys., 25, 1349, 1992.

\bibitem{Bro92}
J.B. Camp, T.W. Darling, and R.E. Brown.
\newblock New Journal of Physics, 71, 783, 1992.

\bibitem{Har13}
A.~H\"arter, A.~Kr\"ukow, M.~Dei\ss, B.~Drews, E.~Tiemann, and
  J.~Hecker~Denschlag.
\newblock Nature Physics, 9, 512--517, 2013.

\bibitem{Har13b}
A.~H\"arter, A.~Kr\"ukow, A.~Brunner, and J.~Hecker~Denschlag.
\newblock Applied Physics Letters, 102, 221115, 2013.

\bibitem{Smi12}
S.~Schmid, A.~H\"arter, A.~Frisch, S.~Hoinka, and J.~Hecker~Denschlag.
\newblock Rev. Sci. Instrum., 83, 053108, 2012.

\end{thebibliography}

\end{document}